\documentclass[preprint,aps,showpacs,nofootinbib,preprintnumbers,amsmath,amssymb]{revtex4-1}
\usepackage{epsfig}
\usepackage{subfigure}
\usepackage{dcolumn}
\usepackage{bm}
\usepackage[usenames ,dvipsnames]{xcolor}
\usepackage{slashed}
\usepackage{float}
\usepackage{graphicx,color}
\usepackage{amsmath}
\usepackage{nonfloat}
\begin{document}
\title{ Semileptonic decays of anti-triplet charmed baryons }

\author{Chao-Qiang Geng$^{1,2,3}$, Chia-Wei Liu$^{2}$, Tien-Hsueh Tsai$^{2}$ and   Shu-Wei Yeh$^{2}$}
\affiliation{
$^{1}$Chongqing University of Posts \& Telecommunications, Chongqing 400065\\
$^{2}$Department of Physics, National Tsing Hua University, Hsinchu 300\\
$^{3}$Physics Division, National Center for Theoretical Sciences, Hsinchu 300}
\date{\today}
\begin{abstract}
We study  the semileptonic decays ${\bf B_c} \to {\bf B_n} \ell^+ \nu_{\ell}$  where ${\bf B_{c(n)}}$ is the anti-triplet-charmed (octet) baryon  
with the $SU(3)_f$ flavor symmetry  and  helicity formalism. In particular, we present  the decay branching ratios of $ {\bf B_c}  \to {\bf B_n}\ell^+ \nu_\ell$ 
in three scenarios: 
(a) an exact $SU(3)_f$ symmetry with equal masses for the anti-triplet-charmed (octet) baryon states of ${\bf B_c}$ (${\bf B_n}$),
(b) $SU(3)_f$  parameters without  the baryonic momentum-transfer dependence, and 
(c)  $SU(3)_f$ with baryonic transition form factors in the heavy quark limit. 
We show that our results are all consistent with the existing data. Explicitly, we  predict that 
${\cal B}( \Xi_c^+ \to \Xi^0 e^+ \nu_{e})=(11.9\pm1.3, 9.8\pm 1.1, 10.7\pm 0.9)\times 10^{-2}$ and
${\cal B}( \Xi_c^0 \to \Xi^- e^+ \nu_{e})=(3.0\pm0.3, 2.4\pm 0.3, 2.7\pm 0.2)\times 10^{-2}$ in the scenarios (a), (b) and (c) agree with
the data of $(14.0^{+8.3}_{-8.7})\times 10^{-2}$ and $(5.6\pm2.6)\times 10^{-2}$ from the CLEO Collaboration, respectively.
In addition, we obtain that ${\cal B}(\Lambda_c^+\to n e^+ \nu_e)=(2.8\pm 0.4, 4.9\pm0.4, 5.1\pm0.4)\times 10^{-3}$ in (a), (b) and (c).
We also examine the  longitudinal asymmetry parameters of  $\alpha({\bf B_c} \to {\bf B_n} \ell^+ \nu_{\ell})$, which are sensitive to the different
scenarios with $SU(3)_f$. Some of the decay branching ratios and asymmetries can be observed by the ongoing experiments at BESIII and LHCb 
as well as the future searches by BELLEII. 
\end{abstract}
\maketitle

\section{introduction}
Very recently, the absolute branching ratio of ${\cal B} (\Xi_c^0 \to \Xi^- \pi^+)=(1.8\pm0.5 )\times 10^{-2}$ has been measured for the first time  by the
Belle collaboration~\cite{Li:2018qak}, which is the golden mode in $\Xi_c^0$ decays. 
In fact, there have been  significant experimental progresses  in observing  weak decays of  charmed baryons~\cite{Tanabashi:2018oca}. 
It is no doubt that we are witnessing a new era of charm physics. 
On the other hand, theoretical studies of charmed baryon decays have faced many difficulties. For instance,
the complicated structures of these baryons with large non-perturbative effects of the quantum chromodynamic (QCD) make us impossible to reliably calculate 
their decays amplitudes from first principles. 
Fortunately, there is a very powerful tool  to explore the charmed baryon decays based on the flavor symmetry of $SU(3)_f$ in the quark model,
which is a model independent way to connect various  decay channels.
Recently, 
 several theoretical  analyses of two and three-body non-leptonic processes of charmed baryons
 have been performed in the literature~\cite{Wang:2017gxe,Geng:2017mxn,Geng:2018bow,Geng:2017esc,Geng:2018plk,Geng:2018rse,Geng:2018upx,Lu:2016ogy}
 based on the newly measured decay branching fractions. 
In particular, the prediction of ${\cal B}(\Xi_c^0 \to \Xi^- \pi^+)=(1.6\pm0.1 )\times 10^{-2}$ from the  $SU(3)_f$ approach~\cite{Geng:2018plk}
is consistent with the measurement by  Belle~\cite{Li:2018qak}.
As a result, we are confidence that the use of $SU(3)_f$ is a good method to examine the weak decays of charmed baryons. 

It is known that the semileptonic decays of   charmed baryons are the cleanest processes as they can be calculated through the QCD factorization approach.
Hence,  
these decays are good platforms to test $SU(3)_f$ and identify the  corresponding breaking effects.
Besides the total branching ratios of these semileptonic decays, the 
 angular distribution asymmetries, which  contain the information of the underlined  dynamics of the decays, 
 can also  constrain the theoretical QCD models. 
In Refs.~\cite{Korner:1994nh,Korner:1991ph,Bialas:1992ny,Kadeer:2005aq},
the helicity formalism has been used to analyze the angular properties of both parent and daughter baryons
  in the   baryonic decays. 
However, theoretical considerations for the asymmetrical parameters in the semileptonic  decays of  charmed baryons with the $SU(3)_f$ symmetry 
have not been systematically examined yet even though they could be  well measured  experimentally. 
Currently, there are only three  experimental data for the semileptonic decays of $\Lambda_c$, given by~\cite{Tanabashi:2018oca} 
\begin{eqnarray}
\label{eq1}
{\cal B}(\Lambda_c^+ \to \Lambda e^+ \nu_{e})&=&(3.6\pm 0.4)\times 10^{-2}\,, \nonumber \\
{\cal B}(\Lambda_c^+ \to \Lambda \mu^+ \nu_{\mu})&=&(3.5 \pm 0.5 )\times 10^{-2}\,, \nonumber \\
\langle\alpha\rangle(\Lambda_c^+ \to \Lambda e^+ \nu_{e})&=&-0.86\pm 0.04 \,,
\end{eqnarray}
where $\alpha$ is the longitudinal asymmetrical parameter. 
The decay of $\Lambda_c^+ \to \Lambda e^+ \nu_{e}$ has been extensively studied in  the literature. 
 In particular, 
 its decay branching ratio has been found to be $ (1.42,1.63, 2.78,2.96,3.6,3.8)\times 10^{-2}$
  by  the heavy quark effective theory (HQET) along with the non-relativistic quark model (NRQM)~\cite{Cheng:1995fe}, QCD light front (LF) approach~\cite{Zhao:2018zcb}, covariant quark model (CQM)~\cite{Gutsche:2014zna,Gutsche:2015rrt},  MIT bag model (MBM)~\cite{PerezMarcial:1989yh}, NRQM~\cite{PerezMarcial:1989yh}
and lattice QCD (LQCD)~\cite{Meinel:2016dqj,Meinel:2017ggx}. Clearly, the  predicted values in the  models except NRQM and LQCD
are inconsistent with the experimental data in Eq.~(\ref{eq1}).
The $SU(3)_f$ structures for the semileptonic decay amplitudes of charmed baryons  are quiet simple. 
In fact, all decay branching ratios and asymmetries  are related by one $SU(3)_f$ parameter, which can be determined by the experimental data
in Eq.~(\ref{eq1})~\cite{Lu:2016ogy,Geng:2017mxn}.
In this study, besides imposing  $SU(3)_f$  in the decay amplitudes~\cite{Geng:2017mxn,Geng:2018bow,Geng:2017esc,Geng:2018plk,Geng:2018rse,Geng:2018upx,Lu:2016ogy,Wang:2017gxe}, we also consider $SU(3)_f$ to be held in each baryonic transition form factor,
which describes the non-perturbative QCD effect in the decay processes to include the mass effect in phase space integration. 
We will systematically examine all semileptonic decays of the anti-triplet charmed baryons.

This paper is organized as follows. In Sec. II, we  write down  decay widths and asymmetrical parameters in terms of 
the helicity formalism.   In Sec. III, after  including the mass corrections in the phase space integrations and imposing the spin symmetry 
in the heavy quark limit to reduce the  $SU(3)_f$ parameters, we show our numerical results.
We present the conclusions in Sec.~IV.

\section{Decay branching ratios and asymmetries}
We concentrate on  the semileptonic decays of ${\bf B_{c}}\to  {\bf B_{n}} \ell^+ \nu_\ell$,
where ${\bf B_c}$ and ${\bf B_n}$ are anti-triplet charmed and  octet  baryon states under the $SU(3)_f$ flavor symmetry, 
defined by
\begin{eqnarray}
{\bf B_c}&=&(\Xi_c^0,-\Xi_c^+,\Lambda_c)\,, \nonumber \\
{\bf B}_n&=&\left(\begin{array}{ccc}
\frac{1}{\sqrt{6}}\Lambda+\frac{1}{\sqrt{2}}\Sigma^0 & \Sigma^+ & p \\
\Sigma^- &\frac{1}{\sqrt{6}}\Lambda-\frac{1}{\sqrt{2}}\Sigma^0  & n \\
\Xi^- & \Xi^0 &-\sqrt{\frac{2}{3}}\Lambda
\end{array}\right)\,,
\end{eqnarray}
while $\ell=e,\mu$
and $\nu_\ell$ are the charged and neutral leptons, respectively.
The decay transition amplitudes are written as
\begin{eqnarray}
{\cal A}({\bf B_{c}}\to  {\bf B_{n}} \ell^+ \nu_\ell)&=&\frac{G_F}{\sqrt{2}}V_{cq}\langle  {\bf B_{n}}|J_{\mu }^{(V-A)}|{\bf B_{c}}\rangle \bar{u}_{\nu}\gamma^{\mu}\frac{(1-\gamma_5)}{2}v_\ell \,,
\end{eqnarray}
where $J_{\mu }^{(V-A)}={\bar q}\gamma_{\mu}(1-\gamma_5)/2c$,
$q=d,s$, $\bar{u}_{\nu}$ and $v_\ell$ are Dirac bispinors, $G_F$ is Fermi constant, and $V_{cq}$ are the
Cabibbo-Kobayashi-Maskawa (CKM) quark mixing matrix elements. 

Since the lepton part can be traced back to the off-shell W-boson ($W_{os}$), 
we can introduce a set of helicity vectors $\epsilon^{\mu}(\lambda_{W})$  in the ${\bf B_c}$ rest frame, given by
\begin{eqnarray}
\epsilon^{\mu}(t)=\frac{1}{\sqrt{q^2}}(q_0,0,0,-p) \nonumber\\
\epsilon^{\mu}( \pm 1)=\frac{1}{\sqrt{2}}(0,\pm 1,-i,0)  \nonumber \\
\epsilon^{\mu}(0)=\frac{1}{\sqrt{q^2}}(p,0,0,-q_0)
\end{eqnarray}
where $\lambda_W=(\pm 1,0)$ and $\lambda_W=t$, 
denoting $J=1$ and $0$ parts of $W_{os}$, respectively, and $q^{\mu}=(q_0;,0,0,p)$ is the four momentum of $W_{os}$ 
with $m_{\ell}^2<q^2<(M_{\bf B_c}-M_{\bf B_n})^2$ and 
\begin{eqnarray}
p&=&\frac{1}{2M_{\bf B_c}}\sqrt{Q_+Q_-}\,, \nonumber \label{pcm}\\
Q_\pm&=&(M_{\bf B_c}\pm M_{\bf B_n})^2-q^2 \,,
\end{eqnarray}
with $M_{\bf B_{c(n)}}$ being the mass of ${\bf B_{c(n)}}$.
Hence, we can decompose the total transition amplitudes into helicity ones~\cite{Kadeer:2005aq}, given by
\begin{eqnarray}
{\cal A}({\bf B_{c}}\to  {\bf B_{n}} \ell^+ \nu_\ell)&=&\sum_{\lambda_{W},\lambda'_{W}=\pm 1,0 ,t}\frac{G_F}{\sqrt{2}}V_{cq} H_{\lambda_2 \lambda_{W}} \bar{u}_{\nu}\gamma_{\beta}\frac{(1-\gamma_5)}{2}v_\ell \epsilon^{*\beta}(\lambda'_{W})g_{\lambda_{W}\lambda'_{W}}\,, \nonumber \\
H_{\lambda_2 \lambda_{W}}&=&H^{V}_{\lambda_2 \lambda_{W}}-H^{A}_{\lambda_2 \lambda_{W}}\,,\ \quad 
H^{V(A)}_{\lambda_2 \lambda_{W}}=\langle  {\bf B_{n}}|J_{\mu }^{V(A)}|{\bf B_{c}}\rangle \epsilon^{\mu}(\lambda_{W}) \,,
\end{eqnarray}
where $ \lambda_{2(W)}$ corresponds to the helicity of the daughter baryon ($W_{os}$)
Note that  the helicity of the parent baryon is fixed by $\lambda_1=\lambda_2-\lambda_{W}$, while $H^{V(A)}_{\lambda_2 \lambda_{W}}=(-)H^{V(A)}_{-\lambda_2 -\lambda_{W}}$ under the parity transformation.
We can also write the helicity amplitudes in terms of the invariant baryonic transition form factors~\cite{Kadeer:2005aq}, 
given by
\begin{eqnarray}
\label{HFF}
H^{V}_{\frac{1}{2}1}&=&\sqrt{2Q_-}\left(-F^V_1-\frac{M_{\bf B_c}+M_{\bf B_n}}{M_{\bf B_c}}F^V_2 \right)\,, \nonumber \\
H^{V}_{\frac{1}{2}0}&=&\frac{\sqrt{Q_-}}{\sqrt{q^2}}\left((M_{\bf B_c}+M_{\bf B_n})F^V_1+\frac{q^2}{M_{\bf B_c}}F^V_2\right)\,,\nonumber \\
H^{V}_{\frac{1}{2}t}&=&\frac{\sqrt{Q_+}}{\sqrt{q^2}}\left((M_{\bf B_c}-M_{\bf B_n})F^V_1+\frac{q^2}{M_{\bf B_c}}F^V_3\right)\,,\nonumber \\
H^{A}_{\frac{1}{2}1}&=&\sqrt{2Q_+}\left(F^A_1-\frac{M_{\bf B_c}-M_{\bf B_n}}{M_{\bf B_c}}F^A_2\right)  \,, \nonumber \\
H^{A}_{\frac{1}{2}0}&=&\frac{\sqrt{Q_+}}{\sqrt{q^2}}\left(-(M_{\bf B_c}-M_{\bf B_n})F^A_1+\frac{q^2}{M_{\bf B_c}}F^A_2\right)\,, \nonumber \\
H^{A}_{\frac{1}{2}t}&=&\frac{\sqrt{Q_-}}{\sqrt{q^2}}\left(-(M_{\bf B_c}+M_{\bf B_n})F^A_1+\frac{q^2}{M_{\bf B_c}}F^A_3\right)\
\end{eqnarray}
 where $F_{1,2,3}^{V,A}$ are defined by \footnote{The formula of invariant form factors used by CLEO~\cite{Hinson:2004pj} can be found in Ref.~\cite{Korner:1991ph}, which have  opposite signs in front of $F_{2}^{V,A}$
	compared with Eq.~(\ref{FF}), so that  there is a sign difference between  our result and that by CLEO for $r=F_2/F_1$.}
\begin{eqnarray}
\label{FF}
 \langle {\bf B_n}|J_\mu^V|{\bf B_c}\rangle & =& \bar{u}_{\bf B_n}(p_{\bf B_n})\bigg[F_1^V(q^2)\gamma_\mu 
 - \frac{F_2^V(q^2)}{M_{\bf B_c}}i\sigma_{\mu \nu}q^\nu+ \frac{F_3^V(q^2)}{M_{\bf B_c}}q_\mu \bigg] u_{\bf B_c}(p_{\bf B_c})\;,\nonumber\\
\langle {\bf B_n}|J_\mu^A|B_{\bf B_c} \rangle &=& \bar{u}_{\bf B_n}(p_{\bf B_n})\bigg[F_1^A(q^2)\gamma_\mu
- \frac{F_2^A(q^2)}{M_{\bf B_c}}i\sigma_{\mu \nu}q^\nu  + \frac{F_3^A(q^2)}{M_{\bf B_c}}q_\mu \bigg]\gamma_5 u_{\bf B_c}(p_{\bf B_c})\;.
\end{eqnarray}
where the $\sigma_{\mu \nu}=\frac{i}{2}[\gamma_{\mu},\gamma_{\nu}]$.
We now parameterize baryon states and quark operators into SU(3) tensor forms, while the
polarization vectors $\epsilon^{\mu}(\lambda_{W})$ are invariant under $SU(3)_f$.  As a result, 
 the transition operators of  $({\bar q}c)^{V(A)}_{\lambda_W}$ are transformed as an anti-triplet (${\bar 3}$),  denoted as $T({\bar 3})=(0,1,1)$, under $SU(3)_f$. 
 Consequently, the 
 helicity amplitudes can be rewritten as
\begin{eqnarray}
H^{V(A)}_{\lambda_2 \lambda_{W}}=a_{\lambda_2\lambda_{W}}^{V(A)}(q^2)({\bf B_n})^{i}_{j}T^j({\bar 3})({\bf B_c})_i\,.
\end{eqnarray}
With the $SU(3)_f$ symmetry, the connections between
the helicity amplitudes of different channels are presented in Table~\ref{H-amp}. 
Since the baryon transition matrix of $\langle  {\bf B_{n}}|(J_{\mu }^{(V-A)}|{\bf B_{c}}\rangle $ is invariant under the CP transformation, all $SU(3)_f$ parameters $a_{\lambda_{W}}^{V(A)}(q^2)$ with the same helicity quantum number are real up to an overall phase. 
\begin{table}[h]
	\caption{Helicity amplitudes of $ {\bf B_c}  \to {\bf B_n}\ell^+ \nu_\ell$ with $SU(3)_f$. }
	\label{H-amp}
	\begin{tabular}{l|l}
		\hline
		channel& $H_{\lambda_2\lambda_{W}}^{V(A)}$\\
		\hline
		$\Lambda_c^+ \to \Lambda \ell^{+}\nu_{\ell}$&$-\sqrt{\frac{2}{3}}a_{\lambda_2\lambda_{W}}^{V(A)}(q^2)$\\
		\hline
		$\Xi_c^+ \to \Xi^0 \ell^{+}\nu_{\ell}$&$-a_{\lambda_2\lambda_{W}}^{V(A)}(q^2)$\\
		\hline
		$\Xi_c^0 \to \Xi^- \ell^{+}\nu_{\ell}$&$a_{\lambda_2\lambda_{W}}^{V(A)}(q^2)$\\
		\hline
		\hline
		$\Lambda_c^+ \to n \ell^{+}\nu_{\ell}$&$a_{\lambda_2\lambda_{W}}^{V(A)}(q^2)$\\
		\hline
		$\Xi_c^+ \to \Sigma^0 \ell^{+}\nu_{\ell}$&$\sqrt{\frac{1}{2}}a_{\lambda_2\lambda_{W}}^{V(A)}(q^2)$\\
		\hline
		$\Xi_c^+ \to \Lambda \ell^{+}\nu_{\ell}$&$-\sqrt{\frac{1}{6}}a_{\lambda_2\lambda_{W}}^{V(A)}(q^2)$\\
		\hline
		$\Xi_c^0 \to \Sigma^- \ell^{+}\nu_{\ell}$&$a_{\lambda_2\lambda_{W}}^{V(A)}(q^2)$\\
		\hline

	\end{tabular}	
\end{table}

The differential decay widths of the semileptonic processes can be written as analytic forms,
given by~\cite{Korner:1994nh,Kadeer:2005aq}
\begin{eqnarray}
\label{Rate}
	\frac{d\Gamma}{dq^2}&=&\frac{1}{3}\frac{G_F^2}{(2\pi)^3}|V_{qc}|^2\frac{(q^2-m^2_{\ell})^2p}{8M^2_{\bf B_c}q^2}
	\Bigg[\left(1+\frac{m^2_{\ell}}{2q^2}\right)  \nonumber \\
	&&\left(|H_{\frac{1}{2}1}|^2+|H_{-\frac{1}{2}-1}|^2+|H_{\frac{1}{2}0}|^2+|H_{-\frac{1}{2}0}|^2\right)
	+\frac{3m^2_{\ell}}{2q^2}\left(|H_{\frac{1}{2}t}|^2+|H_{-\frac{1}{2}t}|^2\right)\Bigg]\,.
\end{eqnarray} 
In the limit of $m_{\ell}= 0 $,  $H_{\lambda_2 t}^{V(A)}$ has no contribution to the branching ratios due to the helicity suppression by the factor of 
$m^2_{\ell}/M^2_{\bf B_{c}}$ as given by Refs.~\cite{Korner:1994nh,Kadeer:2005aq}.
Since all $H_{\lambda_2\lambda_{W}}^{V(A)}$ for the decays correspond to the same $SU(3)_f$ parameters of $a_{\lambda_2\lambda_{W}}^{V(A)}(q^2)$, 
there are several direct relations for the differential decay widths before integrating over the $q^2$ dependences, given by 
\begin{eqnarray}
d\Gamma(\Lambda_c^+ \to \Lambda \ell^{+}\nu_{\ell})&=&\frac{2 p_{\Lambda}}{3p_{\Xi^0}}d\Gamma(\Xi_c^+ \to \Xi^0 \ell^{+}\nu_{\ell})=\frac{2 p_{\Lambda}}{3p_{\Xi^-}}d\Gamma(\Xi_c^0 \to \Xi_c^- \ell^{+}\nu_{\ell})\,,
  \nonumber\\
d\Gamma(\Lambda_c^+ \to n^0 \ell^{+}\nu_{\ell})&=&\frac{2p_{n}}{p_{\Sigma^0}}d\Gamma(\Xi_c^+ \to \Sigma^0 \ell^{+}\nu_{\ell})=\frac{6p_{n}}{p_{\Lambda}}d\Gamma(\Xi_c^+ \to \Lambda \ell^{+}\nu_{\ell}) \nonumber\\
&=&\frac{p_{n}}{p_{\Sigma^-}}d\Gamma(\Xi_c^0 \to \Sigma^- \ell^{+}\nu_{\ell})\,,
\end{eqnarray}  
for the  $c \to (s,d)$ transitions, respectively,
where the masses of the charmed baryons are set to be equal, $p_{\bf B_n}$ is defined in Eq.~(\ref{pcm}) with different masses for the octet baryon. 
Under the exact $SU(3)_f$ limit, $M_{\Lambda_c}=M_{\Xi_c^+}=M_{\Xi_c^0}$ and
$M_p=M_n=M_{\Lambda}=M_{\Sigma^0}=M_{\Sigma^\pm}=M_{\Xi^-}=M_{\Xi^0}$
so that
$p_{\bf B_n}/p_{\bf B'_n}=1$ along with the same  phase space volume. 
Hence, without knowing the $q^2$ dependences of $a_{\lambda_2\lambda_{W}}^{V(A)}(q^2)$, 
we can use one experimental data to derive all other branching ratios.

In the semileptonic decay of ${\bf B_{c}}\to  {\bf B_{n}} \ell^+ \nu_\ell$,  one can write
the asymmetrical parameter $\alpha({\bf B_{c}}\to  {\bf B_{n}} \ell^+ \nu_\ell)$~\cite{Korner:1994nh,Korner:1991ph}, 
which is also known as the longitudinal polarization of the daughter baryon,  defined by
\begin{eqnarray}
 \alpha(q^2) =
 \frac{
(1+\frac{m^2_{\ell}}{2q^2})
\left(|H_{\frac{1}{2}1}|^2-|H_{-\frac{1}{2}-1}|^2+|H_{\frac{1}{2}0}|^2-|H_{-\frac{1}{2}0}|^2\right)
	+\frac{3m^2_{\ell}}{2q^2}\left(|H_{\frac{1}{2}t}|^2-|H_{-\frac{1}{2}t}|^2\right) }
	{	(1+\frac{m^2_{\ell}}{2q^2})
\left(|H_{\frac{1}{2}1}|^2+|H_{-\frac{1}{2}-1}|^2+|H_{\frac{1}{2}0}|^2
	+|H_{-\frac{1}{2}0}|^2\right)+\frac{3m^2_{\ell}}{2q^2}\left(|H_{\frac{1}{2}t}|^2+|H_{-\frac{1}{2}t}|^2\right)} .~
\end{eqnarray} 
Consequently, from Eq.~(\ref{Rate}) one can also define the integrated (averaged) asymmetry    by~\cite{Gutsche:2015rrt,Faustov:2016yza}
{\scriptsize
\begin{eqnarray}
\langle\alpha\rangle &=&
\frac{\int dq^2 \frac{(q^2-m^2_{\ell})^2p}{8M^2_{\bf B_c}q^2} \left[
(1+\frac{m^2_{\ell}}{2q^2})\left(|H_{\frac{1}{2}1}|^2-|H_{-\frac{1}{2}-1}|^2+|H_{\frac{1}{2}0}|^2-|H_{-\frac{1}{2}0}|^2\right)
	+\frac{3m^2_{\ell}}{2q^2}\left(|H_{\frac{1}{2}t}|^2-|H_{-\frac{1}{2}t}|^2\right)\right] }
	{ \int dq^2\frac{(q^2-m^2_{\ell})^2p}{8M^2_{\bf B_c}q^2}\left[
(1+\frac{m^2_{\ell}}{2q^2})\left(|H_{\frac{1}{2}1}|^2+|H_{-\frac{1}{2}-1}|^2+|H_{\frac{1}{2}0}|^2
	+|H_{-\frac{1}{2}0}|^2\right)+\frac{3m^2_{\ell}}{2q^2}\left(|H_{\frac{1}{2}t}|^2+|H_{-\frac{1}{2}t}|^2\right)\right]} \,.
\end{eqnarray} }
If $SU(3)_f$ is an exact symmetry, 
the parameter of $\alpha$ is equal for all decay modes. Clearly, it is a good observable to test  the $SU(3)_f$ flavor symmetry.

\section{Numerical Results}
We first show our numerical results  based on the exact $SU(3)_f$ flavor symmetry with the same mass for all anti-triplet charmed (octet) baryon states
in the second column of Table~\ref{Br}. 
We then  consider the mass effects from the $q^2$ integrations, in which the helicity amplitudes still preserve $SU(3)_f$. 
We treat the $SU(3)_f$ parameters as constants without the $q^2$ dependences, i.e. 
$a_{\lambda_{W}}^{V(A)}(q^2)=$ constant, and present the decay branching ratios in the third column of Table~\ref{Br}.

\begin{table}[!t]
	\caption{Decay branching ratios of $ {\bf B_c}  \to {\bf B_n}\ell^+ \nu_\ell$  with $SU(3)_f$ for (a) equal masses of  ${\bf B_c}$ (${\bf B_n}$),
	(b) $a_{\lambda_{W}}^{V(A)}(q^2)=$ constant and (c) the  baryonic transition form factors in the heavy quark limit.}
	\label{Br}
{	\footnotesize
	\begin{tabular}{c|c|c|c|c|c|c|c|c}
		\hline
Branching ratio&$SU(3)_f$& $SU(3)_f$& $SU(3)_f$& HQET&LF&MBM(NRQM) &LQCD&Data \\
&(a)&(b) & (c) &\cite{Cheng:1995fe}&\cite{Zhao:2018zcb}&\cite{PerezMarcial:1989yh}&\cite{Meinel:2016dqj,Meinel:2017ggx} &
\cite{Tanabashi:2018oca,Li:2018qak,Alexander:1994hp} \\
		\hline
		$10^{2}{\cal B}(\Lambda_c^+ \to \Lambda e^{+}\nu_{e})$&$3.6\pm0.4$&$3.6\pm0.4$&$3.2\pm0.3$&1.42&$1.63$&$2.96(3.60)$
		&$3.80\pm0.22$&$3.6\pm 0.4$\\
		\hline
		$10^{2}{\cal B}(\Lambda_c^+ \to \Lambda \mu^{+}\nu_{\mu})$&$3.5\pm 0.5$&$3.6\pm0.4$&$3.2\pm0.3$&-&-&-
		&$3.69\pm0.22$&$3.5\pm 0.5$\\
        \hline
        $10^{2}{\cal B}(\Xi_c^+ \to \Xi^0 e^{+}\nu_{e})$&$11.9\pm 1.3$&$9.8\pm1.1$&$10.7\pm0.9$&-&$5.39$&1.33(1.01)&-&$14.0^{+8.3}_{-8.7}$\\
		\hline
		$10^{2}{\cal B}(\Xi_c^+ \to \Xi^0 \mu^{+}\nu_{\mu})$&$11.6\pm1.7$&$9.8 \pm 1.1$&$10.8\pm 0.9$&-&-&-&-&-\\
		\hline
		$10^{2}{\cal B}(\Xi_c^0 \to \Xi^- e^{+}\nu_{e})$&$3.0\pm 0.3$&$2.4\pm0.3$&$2.7\pm0.2$&0.86&$1.35$&0.40(0.30)&-&$5.6\pm2.6$\\
		\hline
		$10^{2}{\cal B}(\Xi_c^0 \to \Xi^- \mu^{+}\nu_{\mu})$&$2.9\pm 0.4$&$2.4\pm0.3$&$2.7\pm0.2$&-&-&-&-&-\\
		\hline
		\hline
		$10^{3}{\cal B}(\Lambda_c^+ \to n e^{+}\nu_{e})$ &$2.8\pm 0.4$&$4.9\pm0.5$&$5.1\pm0.4$&-&$2.01$&2.20(3.40)
		&$4.10\pm0.29$&--\\
		\hline
		$10^{3}{\cal B}(\Xi_c^+ \to \Sigma^0 e^{+}\nu_{e})$&$3.1\pm0.4$&$3.8\pm0.4$&$4.6\pm0.4$&-&$1.87$&4.42(4.42)&-&--\\
		\hline
		$10^{4}{\cal B}(\Xi_c^+ \to \Lambda e^{+}\nu_{e})$&$10.3\pm1.5$&$16.6\pm1.8$&$21.8\pm1.8$&-&$8.22$&8.84(8.84)&-&--\\
		\hline
		$10^{4}{\cal B}(\Xi_c^0 \to \Sigma^- e^{+}\nu_{e})$&$15.7\pm2.2$&$19.2\pm2.1$&$23.2\pm1.9$&-&$9.47$&2.24(1.12)&-&--\\
		\hline	
	\end{tabular}	
}
\end{table}

 In order to get more precise numerical values, we use the relations between the helicity amplitudes and invariant form factors in Eq.~(\ref{HFF}).
 Since these form factors have the same relations as those in the exact $SU(3)_f$ limit, they also preserve the $SU(3)_f$ symmetry.
  When we treat the charm quark to be much heaver than other quarks in ${\bf B_c}$, we can apply the spin symmetry in this
  heavy quark limit (HQL) to derive  $F_{i}^{V}(q^2)=F_{i}^{A}(q^2)=F_i(q^2)$ in the ${\bf B_c}\to {\bf B_n}$ transitions~\cite{Korner:1991ph,Mannel:1990vg}. 
  In the following discussions, we will take the HQL in our numerical calculations.
To illustrate our results, we choose the dipole behavior for $F_i(q^2)$ as used by CLED in Ref.~\cite{Hinson:2004pj}, given by
\begin{eqnarray}
F_i(q^2)={F_i\over (1-q^2/M_V^2)^2}\,, 
\end{eqnarray}
where $F_i\equiv F_i(q^2=0)$ and $M_{V}=2.061\hspace{1ex}GeV$, which is the average mass of the lowest excited $D$ and $D_s$ mesons 
with quantum numbers of $J^p=1^{-}$ .
By using the experimental data of $\langle\alpha\rangle(\Lambda_c^+ \to \Lambda e^{+}\nu_{e})=-0.86\pm0.04$ in Eq.~(\ref{eq1}), 
we can fix the ratio of $r=F_2/F_1$, which is plotted  in Fig.~1 with two possible solutions.
\begin{figure}[t]
	\label{f1}
	\includegraphics[width=5.5in]{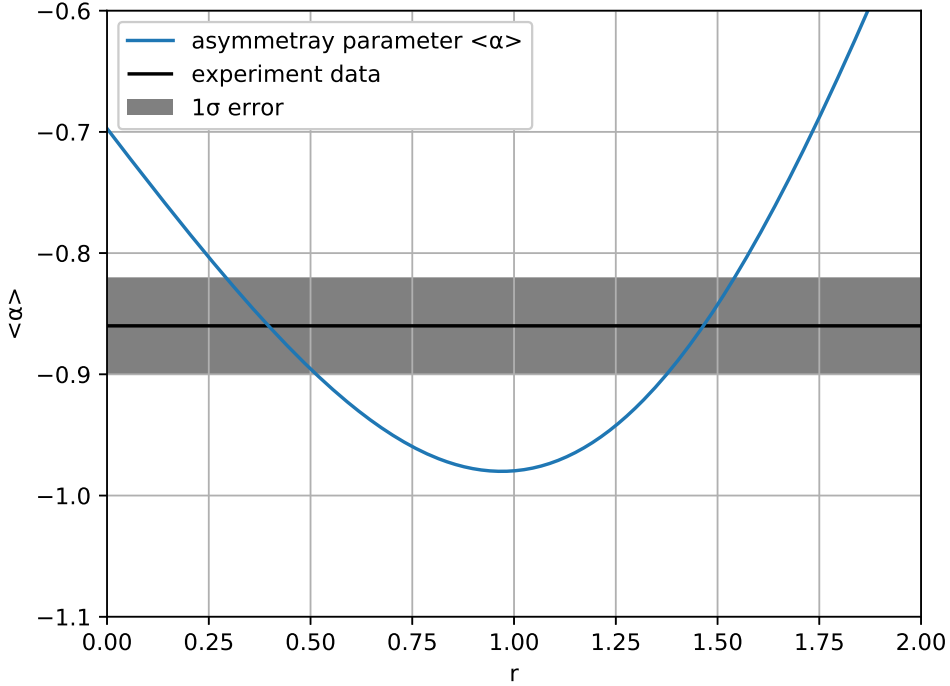}
	\caption{Average asymmetry parameter $<\alpha>$ for $\Lambda_c^+ \to \Lambda e^{+}\nu_{e}$  as a function of $r=F_2/F_1$}. 
\end{figure}
 Since the absolute value of $r$ is expected to be smaller than 1, we select the  solution $r=0.40^{+0.12}_{-0.11}$ . 
 We also perform the minimum $\chi^2$ method to fit $F_1$ and $F_2$ with two measured branching ratios of
 $\Lambda_c \to \Lambda \ell^+ \nu_{\ell}$ ($\ell=e,\mu$) and  one average asymmetry parameter in Eq.~(\ref{eq1})
  along with ${\cal B}(\Xi_c^0 \to \Xi^- e^{+}\nu_{e})=(2.4\pm0.3)\times 10^{-2}$
  predicted by  $SU(3)_f$ given by the first and second columns in Table~\ref{Br}. 
  We obtain that $ (F_1,F_2)=(0.62\pm0.03,0.25\pm0.08) $\footnote{We note that 
  as $F_1$ and $F_2$ are correlated, the correlation coefficient is found to be $0.622$, which will be used to evaluate the errors of our results in the fit.} 
  and $\chi^2/d.o.f=1.2$ with $d.o.f=2$.
 It is interesting to see that the ratio of $r=0.40\pm0.11$
  from the $\chi^2$ fitting is  consistent with the value from the direct calculation.
  Our results for the decay branching ratios with $SU(3)_f$ and the baryonic transition form factors in the HQL 
  are shown in the 4th column of Table~\ref{Br}. 
  
  In Table~\ref{Br}, the data of ${\cal B}(\Xi_c^0 \to \Xi^- e^+ \nu_e)=(5.6\pm2.6)\times 10^{-2}$ is derived from the ratio of  
${\cal B}(\Xi_c^0 \to \Xi^- e^+ \nu_e)/{\cal B}(\Xi_c^0 \to \Xi^- \pi^+)=3.1\pm1.1$ given by 
the CLEO Collaboration~\cite{Alexander:1994hp,Tanabashi:2018oca} and 
the recent measurement of  ${\cal B}(\Xi_c^0 \to \Xi^- \pi^+)=1.8 \pm 0.5$  by Belle~\cite{Li:2018qak},
while that of ${\cal B}(\Xi_c^+ \to \Xi^0 e^+ \nu_e)=(14.0^{+8.3}_{-8.7})\times 10^{-2}$ is resulted from the data
of ${\cal B}(\Xi_c^+ \to \Xi^0 e^+ \nu_e)/{\cal B}(\Xi_c^+ \to \Xi^-\pi^+ \pi^+)=2.3^{+0.7}_{-0.8}$~\cite{Alexander:1994hp,Tanabashi:2018oca} 
and the extracted value  of ${\cal B}(\Xi_c^+ \to \Xi^-\pi^+ \pi^+)=(6.2\pm3.1)\times 10^{-2}$~\cite{Geng:2018upx}
from the data of ${\cal B}(\Xi_c^+ \to \Sigma^-\pi^+ \pi^+)/{\cal B}(\Xi_c^+ \to \Xi^-\pi^+ \pi^+)=0.18\pm0.09$~\cite{Tanabashi:2018oca}.

As shown in Table~\ref{Br}, all of results with $SU(3)_f$ agree well with the existing data.
We note that the predicted value of ${\cal B}(\Lambda_c^+ \to n e^{+}\nu_{e})=(2.93\pm0.34)\times 10^{-3}$ with $SU(3)_f$ in Ref.~\cite{Lu:2016ogy}
corresponds to
our result of $(2.8\pm0.4)\times 10^{-3}$ in Table~\ref{Br}(a) with the exact $SU(3)_f$ symmetry,
while  that of $4.1\times 10^{-3}$ by the LQCD is consistent with our results in Table~\ref{Br}(b) and (c).
In addition, it is interesting to see that,  after multiplying a factor of 2 
 the LF results in Ref.~\cite{Zhao:2018zcb}, which apparently do not agree with the experimental data, 
 almost match up our predicted values in Table~\ref{Br}.
  Moreover, we remark that our value for the inclusive decay branching ratio of  
  ${\cal B}(\Lambda_c^+ \to X e^+ \nu_e)\sim 3.9\times 10^{-2}$
  is  also consistent with the recent data of  
  ${\cal B}(\Lambda_c^+ \to X e^+ \nu_e)=(3.95 \pm 0.35)\times 10^{-2}$ measured by BESIII~\cite{Ablikim:2018woi}.  

  Our results for $\langle\alpha\rangle ({\bf B_c}  \to {\bf B_n}\ell^+ \nu_\ell)$
  based on the $SU(3)_f$ symmetry and  baryonic form factors with the HQL are listed  
   in Table~\ref{result}. 
   From Table~\ref{result}, it is interesting to see that the asymmetry parameters in the muon modes are quite different from
the corresponding electron ones.
We note that in the scenarios  of (a) and (b)
   with $SU(3)_f$, all values of $\langle\alpha\rangle({\bf B_c}  \to {\bf B_n}\ell^+ \nu_\ell)$ are found to be $-0.86\pm0.04$
   fitted by the data of  $\langle\alpha\rangle(\Lambda_c^+ \to \Lambda e^{+}\nu_{e})$ in Eq.~(\ref{eq1})
   because of the cancellation between numerator and denominator integration values. 
\begin{table}[!t]
	\caption{Averaged decay asymmetries of  $ {\bf B_c}  \to {\bf B_n}\ell^+ \nu_\ell$ 
	with $SU(3)_f$ and the baryonic transition form factors in the heavy quark limit. }
	\label{result}
	\begin{tabular}{c|c}
		\hline
	Channel	 & Asymmetry  $\langle \alpha \rangle$  \\
		\hline
$\Lambda_c^+ \to \Lambda \ell^{+}\nu_{\ell}$& $-0.86\pm0.04$\\
		\hline
$\Xi_c^+ \to \Xi^0 \ell^{+}\nu_{\ell}$&$-0.83\pm0.04$\\
		\hline
$\Xi_c^0 \to \Xi^- \ell^{+}\nu_{\ell}$&$-0.83\pm0.04$\\
		\hline
		\hline
$\Lambda_c^+ \to n \ell^{+}\nu_{\ell}$& $-0.89\pm0.04$\\
		\hline
$\Xi_c^+ \to \Sigma^0 \ell^{+}\nu_{\ell}$& $-0.85\pm0.04$\\
		\hline
$\Xi_c^+ \to \Lambda \ell^{+}\nu_{\ell}$& $-0.86\pm0.04$\\
		\hline
$\Xi_c^0 \to \Sigma^- \ell^{+}\nu_{\ell}$& $-0.85\pm0.04$\\
		\hline	
	\end{tabular}	
\end{table}

\section{Conclusions}

We have studied  the semileptonic decays ${\bf B_c} \to {\bf B_n} \ell^+ \nu_{\ell}$  with the $SU(3)_f$ flavor symmetry  and  helicity formalism. 
We have considered the baryonic transition form factors to include the mass effects in the $q^2$ integrations.
In particular, we have concentrated on  three scenarios: 
(a) an exact $SU(3)_f$ symmetry, in which the masses of the anti-triplet-charmed (octet) baryon states of ${\bf B_c}$ (${\bf B_n}$) are equal,
(b) $SU(3)_f$  parameters without  the baryonic momentum-transfer dependence, i.e., $a_{\lambda_{W}}^{V(A)}(q^2)=$ constant,
and 
(c)  $SU(3)_f$ with baryonic transition form factors in the HQL, so that $F_{i}^{V}(q^2)=F_{i}^{A}(q^2)=F_i(q^2)$. 
We have demonstrated  that our results are all consistent with the current data. 
Explicitly, we have found that
${\cal B}( \Xi_c^+ \to \Xi^0 e^+ \nu_{e})=(11.9\pm1.3, 9.8\pm 1.1, 10.7\pm 0.9)\times 10^{-2}$ and
${\cal B}( \Xi_c^0 \to \Xi^- e^+ \nu_{e})=(3.0\pm0.3, 2.4\pm 0.3, 2.7\pm 0.2)\times 10^{-2}$ in the scenarios (a), (b) and (c), 
which are lower than but consistent with the data of $(14.0^{+8.3}_{-8.7})\times 10^{-2}$ and $(5.6\pm2.6)\times 10^{-2}$ from the CLEO Collaboration, respectively.
We have predicted  that ${\cal B}(\Lambda_c^+\to n e^+ \nu_e)=(2.8\pm 0.4, 4.9\pm0.4, 5.2\pm0.4)\times 10^{-3}$ in (a), (b) and (c),
which agree with the previous analysis with $SU(3)_f$ in Ref.~\cite{Lu:2016ogy} as well as that the LQCD~\cite{Meinel:2016dqj,Meinel:2017ggx}.
This mode can be observed at the  experiments by BELLE and BESIII. 

We have also explored the  longitudinal polarization asymmetries in  $\langle\alpha\rangle({\bf B_c} \to {\bf B_n} \ell^+ \nu_{\ell})$.
These asymmetries  are good observables to test $SU(3)_f$ as they are sensitive to the different scenarios. 
We have given that $\langle\alpha\rangle({\bf B_c} \to {\bf B_n} \ell^+ \nu_\ell)$ are around $-0.83\sim -0.89$ 
for both $\ell=(e,\mu)$~\cite{Faustov:2016yza,Gutsche:2015rrt}.

Finally, we remark that the  semileptonic decays of $\Xi_c^{+,0}$ are accessible to not only the current experimental charmed facilities,
but the future ones, such as BELLEII and upgraded BESIII.

\section*{ACKNOWLEDGMENTS}
This work was supported in part by National Center for Theoretical Sciences and 
MoST (MoST-104-2112-M-007-003-MY3 and MoST-107-2119-M-007-013-MY3).


\begin{thebibliography}{99}
	
	

\bibitem{Li:2018qak} 
Y.~B.~Li {\it et al.} [Belle Collaboration],
arXiv:1811.09738 [hep-ex].

\bibitem{Tanabashi:2018oca} 
M.~Tanabashi {\it et al.} [Particle Data Group],
Phys.\ Rev.\ D {\bf 98},  030001 (2018).


\bibitem{Geng:2017mxn}
C.Q.~Geng, Y.K.~Hsiao, C.W.~Liu and T.H.~Tsai,
JHEP {\bf 1711}, 147 (2017). 

\bibitem{Lu:2016ogy} 
C.~D.~Lu, W.~Wang and F.~S.~Yu,
Phys.\ Rev.\ D {\bf 93},  056008 (2016).


\bibitem{Geng:2017esc} 
  C.~Q.~Geng, Y.~K.~Hsiao, Y.~H.~Lin and L.~L.~Liu,
  Phys.\ Lett.\ B {\bf 776}, 265 (2018).


\bibitem{Geng:2018plk}
C.Q.~Geng, Y.K.~Hsiao, C.W.~Liu and T.H.~Tsai,
Phys.\ Rev.\ D {\bf 97}, 073006 (2018).


\bibitem{Geng:2018bow} 
C.Q.~Geng, Y.K.~Hsiao, C.W.~Liu and T.H.~Tsai,
Eur.\ Phys.\ J.\ C {\bf 78}, 593 (2018). 


\bibitem{Geng:2018upx} 
C.~Q.~Geng, Y.~K.~Hsiao, C.~W.~Liu and T.~H.~Tsai,
arXiv:1810.01079 [hep-ph].

\bibitem{Geng:2018rse} 
C.~Q.~Geng, C.~W.~Liu and T.~H.~Tsai,
arXiv:1812.08508 [hep-ph].


\bibitem{Wang:2017gxe} 
D.~Wang, P.~F.~Guo, W.~H.~Long and F.~S.~Yu,
JHEP {\bf 1803}, 066 (2018).

\bibitem{Korner:1991ph} 
J.~G.~Korner and M.~Kramer,
Phys.\ Lett.\ B {\bf 275}, 495 (1992).

\bibitem{Bialas:1992ny} 
P.~Bialas, J.~G.~Korner, M.~Kramer and K.~Zalewski,
Z.\ Phys.\ C {\bf 57}, 115 (1993).

\bibitem{Korner:1994nh} 
J.~G.~Korner, M.~Kramer and D.~Pirjol,
Prog.\ Part.\ Nucl.\ Phys.\  {\bf 33}, 787 (1994).

\bibitem{Kadeer:2005aq} 
A.~Kadeer, J.~G.~Korner and U.~Moosbrugger,
Eur.\ Phys.\ J.\ C {\bf 59}, 27 (2009).

\bibitem{Cheng:1995fe} 
H.~Y.~Cheng and B.~Tseng,
Phys.\ Rev.\ D{\bf 53}, 1457 (1996).

\bibitem{Hinson:2004pj} 
J.~W.~Hinson {\it et al.} [CLEO Collaboration],
Phys.\ Rev.\ Lett.\  {\bf 94}, 191801 (2005)
doi:10.1103/PhysRevLett.94.191801
[hep-ex/0501002].

\bibitem{Zhao:2018zcb} 
Z.~X.~Zhao,
Chin.\ Phys.\ C {\bf 42},  093101 (2018).



\bibitem{Gutsche:2014zna} 
T.~Gutsche, M.~A.~Ivanov, J.~G.~Korner, V.~E.~Lyubovitskij and P.~Santorelli,
Phys.\ Rev.\ D{\bf 90},  114033 (2014).

\bibitem{Gutsche:2015rrt} 
T.~Gutsche, M.~A.~Ivanov, J.~G.~Korner, V.~E.~Lyubovitskij and P.~Santorelli,
Phys.\ Rev.\ D{\bf 93}, 034008 (2016).

\bibitem{PerezMarcial:1989yh} 
R.~Perez-Marcial, R.~Huerta, A.~Garcia and M.~Avila-Aoki,
Phys.\ Rev.\ D{\bf 40}, 2955 (1989).

\bibitem{Faustov:2016yza} 
R.~N.~Faustov and V.~O.~Galkin,
Eur.\ Phys.\ J.\ C {\bf 76}, no. 11, 628 (2016)

\bibitem{Meinel:2017ggx} 
S.~Meinel,
Phys.\ Rev.\ D{\bf 97}, 034511 (2018).
\bibitem{Meinel:2016dqj} 
S.~Meinel,
Phys.\ Rev.\ Lett.\  {\bf 118}, 082001 (2017).

\bibitem{Mannel:1990vg} 
T.~Mannel, W.~Roberts and Z.~Ryzak,
Nucl.\ Phys.\ B {\bf 355}, 38 (1991).

\bibitem{Alexander:1994hp} 
  J.~P.~Alexander {\it et al.} [CLEO Collaboration],
  Phys.\ Rev.\ Lett.\  {\bf 74}, 3113 (1995)
  Erratum: [Phys.\ Rev.\ Lett.\  {\bf 75}, 4155 (1995)].

\bibitem{Ablikim:2018woi} 
M.~Ablikim {\it et al.} [BESIII Collaboration],
Phys.\ Rev.\ Lett.\  {\bf 121},  251801 (2018).
\end{thebibliography}
\end{document}